# Fast and Robust High-Dimensional Sparse Representation Recovery Using Generalized SL0

Milad Nazari, Ali Mehrpooya, Zahra Abbasi, Mehdi Nayebi, M.Hassan Bastani

Faculty of Electrical Engineering

Sharif University of Technology

*Abstract*- Sparse representation can be described in high dimensions and used in many applications, including MRI imaging and radar imaging. In some cases, methods have been proposed to solve the high-dimensional sparse representation problem, but main solution is converting high-dimensional problem into one-dimension. Solving the equivalent problem had very high computational complexity. In this paper, the problem of high-dimensional sparse representation is formulated generally based on the theory of tensors, and a method for solving it based on SL0 (Smoothed Least zero-nor) is presented. Also, the uniqueness conditions for solution of the problem are considered in the high-dimensions. At the end of the paper, some numerical experiments are performed to evaluate the efficiency of the proposed algorithm and the results are presented.

## I. Introduction

Sparse representation and sparse modeling is a topic that has attracted many researchers in the last two decades [1, 2]. The sparse representation problem is an underdetermined linear equation system in which the number of unknowns are greater than the number of equations [3]. The number of zero elements in unknown vector is assumed much larger than non-zeros unknown that means the unknown vector is sparse. The sparse recovery is a method to solve this kind of equation systems based on solving an optimization problem. There are many applications for sparse representation such as compressed sensing [1, 3], blind source separation [4, 5], telecommunication and channel estimation [6, 7], radar imaging [8, 9], machine vision [10, 11]. The compressed sensing might be considered as a turning point for sparse representation.

Using compressed sensing (CS), the sparse signals can be reconstructed at sampling rates below the Nyquist rate by solving a sparse representation problem [1]. Compression sensing, which is based on sparse representation, has been used in various applications such as MRI imaging and radar imaging. M. Lustig [12] proposed a method for rapid MRI imaging using compressed sensing in which the sampling time decreases several times without any significant deterioration in MRI images. Since MRI imaging can be considered inherently three-dimensional, high-dimensional compressed sensing can be used to further increase the speed and accuracy of imaging. Also, a 3-dimensional sensing can also be used in other applications such as radar imaging. Yu-Fei Gao et al. [7] developed a method based on Tucker decomposition for 3-D SAR imaging. However, in most cases when dealing with a sparse problem in 2 or 3 dimensions, the problem is converted to one-dimension and be solved by conventional methods which is not optimal in terms of time and amount of computing.

In this paper a new SL0-based method (Smoothed Least zero-norm) is proposed to solve high-dimensional sparse representation problem. The Organization of this paper is as follow: In Section II, the mathematical model of high-dimensional sparse representation problem is expressed. The uniqueness theorem is generalized for high-dimensional problem in Section III. In Section IV, proposed method for solving high-dimensional problem is explained. Some numerical simulations are presented in Section V to evaluate the efficiency of proposed method. Finally, the conclusion of this paper is expressed in section VI.

# II. Methodology

## *A. Conventional Sparse Representation Problem*

Consider a linear equation system with a number of unknown variables greater than the number of equations. In other words, in the equation system

$$\mathrm{y} = \mathrm{A}\mathrm{x} \tag{1}$$

where $\mathrm{x} \in \mathbb{R}^{n\times 1}$ is unknown variables vector, $\mathrm{y} \in \mathbb{R}^{m\times 1}$ is information vector and $\mathrm{A} \in \mathbb{R}^{m\times n}$ is the coefficients matrix, the number of columns in the matrix $\mathrm{A}$ is greater than the number of rows. In this case, $\mathrm{A}$ called over complete Dictionary Matrix and equation system is underdetermined which has an unlimited number of solutions. The number of columns $\mathrm{A}$ is equal to the number of elements in $\mathrm{x}$ (number of unknown variables) and the number of rows is the same number of elements in y (number of equations). The Eq.(1) can be rewritten as

$$\mathrm{y} = \mathrm{a}_1 x_1 + \mathrm{a}_2 x_2 + \cdots + \mathrm{a}_n x_n = \sum_{k=1}^{n} x_k \mathrm{a}_k \tag{2}$$

where $\mathrm{a}_k \in \mathbb{R}^{m\times 1}$ for $\forall k = 1,2,\dots,n$ are columns of dictionary matrix and called dictionary atoms.

Donoho and Elad in [13] showed that the underdetermined system would have unique solution under certain conditions. They have proven that if the x is sparse enough, which means that most of its elements be zero, the solution of sparse representation problem will be unique. So the unique solution of sparse representation problem can be obtained by solving an optimization problem as

$$\min_{\mathrm{x}} \|\mathrm{x}\|_0 \qquad \text{subject to} \quad \mathrm{y} = \mathrm{Ax} \tag{3}$$

where $\|\mathrm{x}\|_0$ is zero norm of x and is equal to number of nonzero elements of it. The above problem is unstable that means the small noise or numerical error in y lead to significant change in solution for x [14]. The robust version of sparse representation problem can be written as

$$\min_{\mathrm{x}} \|\mathrm{x}\|_0 \qquad \text{subject to} \quad \|\mathrm{y} - \mathrm{Ax}\|_2^2 \le \varepsilon \tag{4}$$

where $\varepsilon$ is noise (or error) energy and denotes the maximum acceptable energy of error in equation $\mathrm{y} = \mathrm{Ax}$.

## *B. High-Dimensional Sparse Representation Problem*

In general, the vectors $\mathrm{x} \in \mathbb{R}^{n\times 1}$ and $\mathrm{y} \in \mathbb{R}^{m\times 1}$ can be replaced with matrices or in general with tensors as $\mathcal{X} \in \mathbb{R}^{N_1\times N_2 \dots \times N_D}$ and $\mathcal{Y} \in \mathbb{R}^{M_1\times M_2 \dots \times M_D}$ which have *D* dimensions. In this case, the conventional sparse problem, which is in one-dimension, turned to a high-dimensional problem. This problem have $D$ dictionary matrices as $\mathrm{A}^{(d)} \in \mathbb{R}^{M_d\times N_d}$ for $d = 1,2,\dots,D$ which are wide $(M_d < N_d)$. Also each element of tensor $\mathcal{Y}$ can be represented as

$$y_{i_1 i_2 i_3 \dots i_D} = \sum_{j_1=1}^{N_1} \sum_{j_2=1}^{N_2} \cdots \sum_{j_D=1}^{N_D} x_{j_1 j_2 j_3 \dots j_D} {a^{(1)}}_{i_1 j_1} {a^{(2)}}_{i_2 j_2} \dots {a^{(D)}}_{i_D j_D} \tag{5}$$

Moreover, the linear equation system for tensors $\mathcal{X}$ and $\mathcal{Y}$ can be written as

$$\mathcal{Y} = \mathcal{X} \times_1 \mathrm{A}^{(1)} \times_2 \mathrm{A}^{(2)} \dots \times_D \mathrm{A}^{(D)} \tag{6}$$

where $\times_d$ denotes the $d$-mode product can be calculated as

$$\left(\mathcal{X} \times_d \mathrm{A}^{(d)}\right)_{j_1 \dots j_{d-1} i_d j_{d+1} \dots j_D} = \sum_{j_d=1}^{N_d} x_{j_1 \dots j_d \dots j_D} a^{(d)}{}_{i_d j_d} \tag{7}$$

The Eq.(7) can be implicated with a matrix product. If $\mathcal{T} = \mathcal{X} \times_d \mathrm{A}^{(d)}$, then two below multiplications are equal

$$\mathcal{T} = \mathcal{X} \times_d \mathrm{A}^{(d)} \Leftrightarrow \mathrm{T}_{(d)} = \mathrm{A}^{(d)} \mathrm{X}_{(d)} \tag{8}$$

where $\mathrm{X}_{(d)} \in \mathbb{R}^{N_d \times (N_1 \dots N_{d-1} N_{d+1} \dots N_D)}$ is matrix expansion of tensor $\mathcal{X}$ in $d$-mode (also called $d$-mode matricization). For example, suppose $\mathcal{X}$ and $\mathcal{Y}$ be 3-dimensional tensors. The graphical representation of $\mathrm{X}_{(1)}$, $\mathrm{X}_{(2)}$ and $\mathrm{X}_{(3)}$ which are matricization of $\mathcal{X}$ in three different modes, is shown in Fig. 1. In addition,

Fig. 2 represents a graphical demonstration of $\mathcal{Y} = \mathcal{X} \times_1 \mathrm{A}^{(1)} \times_2 \mathrm{A}^{(2)} \times_3 \mathrm{A}^{(3)}$. In this example, the tensors $\mathcal{X}$ and $\mathcal{Y}$ are represented by cube and the dictionary matrices are shown by rectangular.

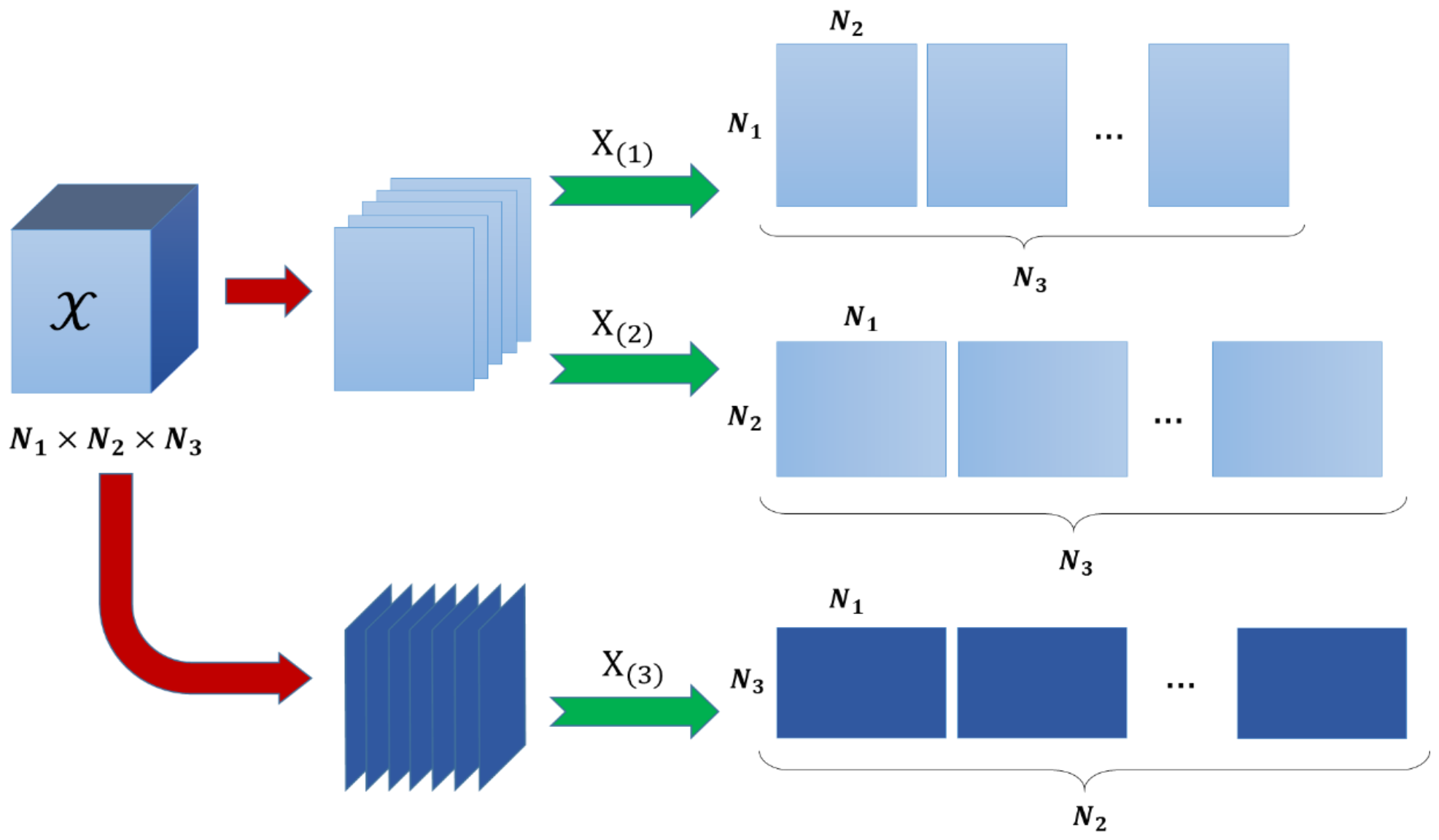

***Fig. 1*** *Matricization (Unfolding) of a 3-dimensional tensor. $X_{(d)}$ is $d$-mode matricization.*

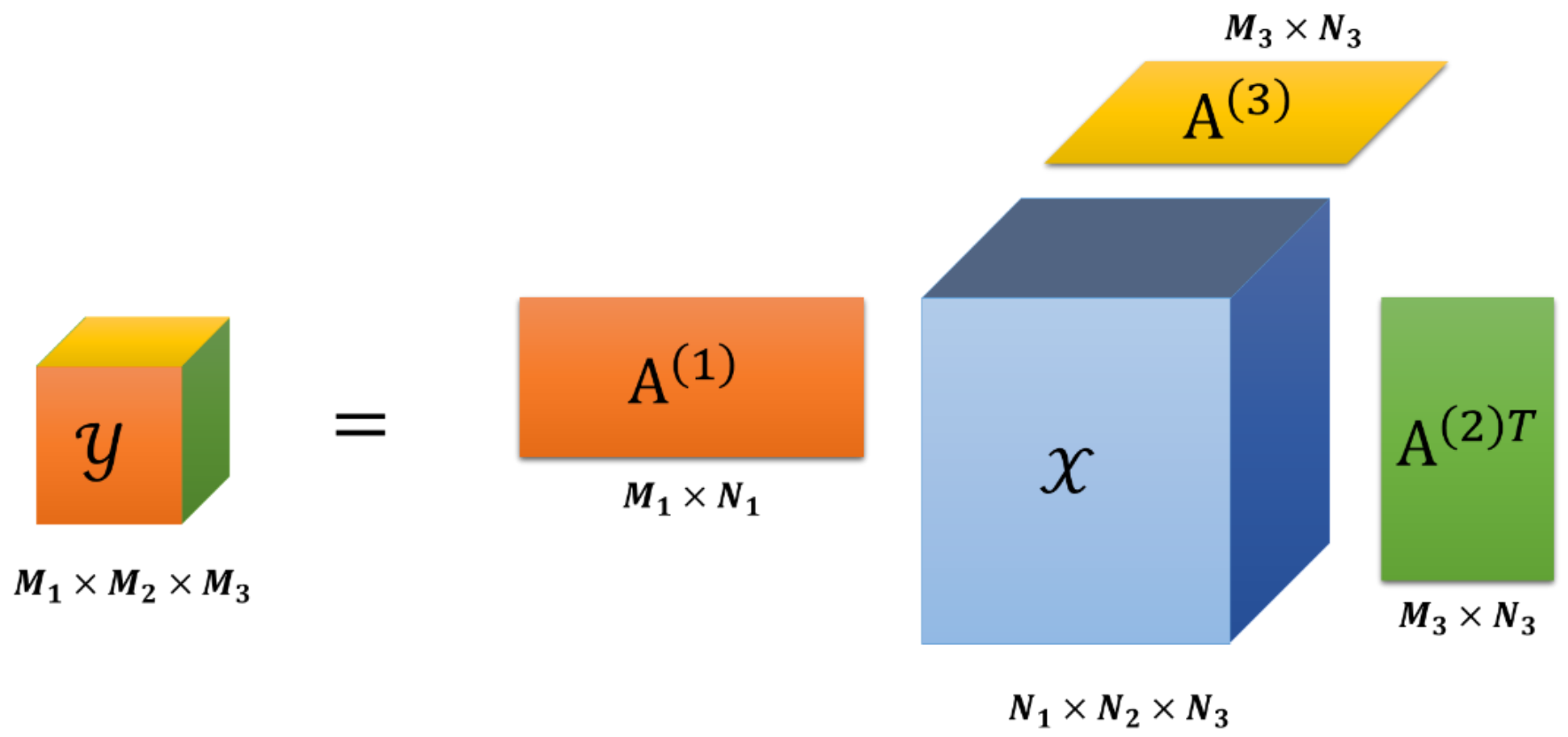

***Fig. 2*** *A graphical demonstration of* $\mathcal{Y} = \mathcal{X} \times_1 A^{(1)} \times_2 A^{(2)} \times_3 A^{(3)}$.

Since the size of tensor $\mathcal{Y}$ is larger than tensor $\mathcal{X}$ in all dimensions, it can be inferred that the tensor equation described in Eq. (6) have an infinite solution for $\mathcal{X}$ as in the conventional case. Also, uniqueness theorem can be expressed here too. It can be claimed that if tensor $\mathcal{X}$ be sparse enough, the solution of the problem will be unique. This claim is expressed and proved in the next section as a theorem. Therefore, high-dimensional sparse representation problem can be written as

$$\min_{\mathcal{X}} \|\mathcal{X}\|_0 \qquad \text{subject to.} \quad \mathcal{Y} = \mathcal{X} \times_1 A^{(1)} \times_2 A^{(2)} \dots \times_D A^{(D)} \tag{9}$$

As in the conventional case, to make the problem in Eq. (9) robust, it could be rewritten as

$$\min_{\mathcal{X}} \|\mathcal{X}\|_0 \qquad \text{subject to.} \quad \left\|\mathcal{Y} - \mathcal{X} \times_1 A^{(1)} \times_2 A^{(2)} \dots \times_D A^{(D)}\right\|_2^2 \leq \varepsilon \tag{10}$$

where $\varepsilon$ is the maximum energy of an acceptable error in the tensor equation descripted in Eq. (6).

### *C. Equivalent 1-Dimensional Sparse Representation Problem*

One of the solutions to the problem of Eq. (9) is to convert the problem of high-dimensional problem into a one-dimensional one and use methods such as SL0 [15], Basis Pursuit [16], Iterative Thresholding [17], etc to solve the conventional sparse representation problem. For this conversion, the tensors $\mathcal{X}$ and $\mathcal{Y}$ should be vectorized as

$$\mathrm{y} = \mathrm{vec}(\mathcal{Y}), \quad \mathrm{x} = \mathrm{vec}(\mathcal{X}) \tag{11}$$

where $\text{vec}(.)$ denotes vectorization operation that by applying it two vector $\mathrm{x} \in \mathbb{R}^{N_1 N_2 \dots N_D \times 1}$ and $\mathrm{y} \in \mathbb{R}^{M_1 M_2 \dots M_D \times 1}$ are achieved. In the next step, the dictionary matrix for 1-dimensional problem (A) is obtained by Kronecker product of matrices $\mathrm{A}^{(1)}, \mathrm{A}^{(2)}, \dots, \mathrm{A}^{(D)}$ as [18]

$$\mathrm{A} = \mathrm{A}^{(1)} \otimes \mathrm{A}^{(2)} \dots \otimes \mathrm{A}^{(D)} \tag{12}$$

where $\otimes$ is Kronecker product and is calculated for two matrix $\mathrm{U} \in \mathbb{C}^{M \times N}$ and $\mathrm{V} \in \mathbb{C}^{P \times Q}$ as

$$\mathrm{W} = \mathrm{U} \otimes \mathrm{V} = \begin{bmatrix} u_{11}v_{11} & u_{11}v_{12} & & & \\ u_{11}v_{21} & u_{11}v_{22} & \cdots & & \\ \vdots & & \ddots & \vdots & \\ & & & u_{MN}v_{PQ} & u_{MN}v_{PQ} \\ & & \cdots & u_{MN}v_{PQ} & u_{MN}v_{PQ} \end{bmatrix} \tag{13}$$

where the size of matrix W is as $MP \times NQ$ [18]. Therefore, the matrix A in Eq. (12) can be expressed as $\mathrm{A} \in \mathbb{R}^{(M_1 M_2 \dots M_D) \times (N_1 N_2 \dots N_D)}$. Finally, the equivalent 1-dimensional problem is achieved as $\mathrm{y} = \mathrm{Ax}$.

Due to the presence of the Kronecker product in calculation A, the size of this matrix is usually too large. Therefore, a small increase in the size of the dictionary matrices $\mathrm{A}^{(1)}, \mathrm{A}^{(2)}, \dots, \mathrm{A}^{(D)}$ or tensors $\mathcal{X}$ and $\mathcal{Y}$ leads to a large increase in the amount and time of computation. So solving the equivalent problem is usually not practicable. For a better understanding of the size of A in equivalent problem, assume $D = 3$ and the size of the matrices $\mathrm{A}^{(1)}, \mathrm{A}^{(2)}, \mathrm{A}^{(3)}$ are $50 \times 100$. So The size of the matrix A will be $50^3 \times 100^3$, which is very large. To solve this problem, in Section IV, we will propose a method to solve the high-dimensional sparse representation problem directly and without converting to 1-dimensional mode. But before that, the uniqueness theorem in high-dimension is expressed in next section.

## III. Uniqueness

In Section II, it was mentioned that if the vector x is sparse enough in the equation $\mathrm{y} = \mathrm{Ax}$, then the solution to the problem is unique. But how much sparse should x be? In this section, the uniqueness theorem is expressed for one-dimensional mode and then generalized for the high dimensions problem.

**Theorem 1** Consider sparse representation problem described in Eq. (3). The solution to problem for the unknown vector $\mathrm{x}^*$ is unique if and only if the following condition holds [13, 19]:

$$\|\mathrm{x}^*\|_0 \leq \frac{\text{Spark}(\mathrm{A})}{2} \tag{14}$$

Where $\mathrm{Spark(A)}$ is the *Spark* property of the matrix $\mathrm{A}$ which is equal to the minimum number of linear dependent columns of $\mathrm{A}$.

Since calculating $\mathrm{Spark(A)}$ is difficult as much as solving the sparse problem, it usually uses another criterion, called *Coherency*, to verify the uniqueness of the solution. The Coherency coefficient of matrix $\mathrm{A}$, shown with $\mu(\mathrm{A})$, is equal to maximum coherency between columns of $\mathrm{A}$ and calculate as

$$\mu(\mathrm{A}) = \max_{i \neq j} \frac{\left|\mathrm{a_i}^T \mathrm{a}_j\right|}{\|\mathrm{a}_i\|_2 \left\|\mathrm{a}_j\right\|_2} \tag{15}$$

where $\mathrm{a}_i$ and $\mathrm{a}_j$ are two non-identical columns of $\mathrm{A}$. For any wide matrix $\mathrm{A}$, the following relation exists between Sparke and Coherency:

$$\mathrm{Spark(A)} \geq 1 + \frac{1}{\mu(\mathrm{A})} \tag{16}$$

Since the calculation of $\mu(\mathrm{A})$ is much easier than $\mathrm{Spark(A)}$, it is possible to state Theorem 2 as another uniqueness theorem, which is weaker than the Theorem 1.

**Theorem 2** Consider the vector $\mathrm{x}^*$ as the solution to sparse representation problem described in Eq. (3). If this vector satisfies the following condition:

$$\|\mathrm{x}^*\|_0 < \frac{1}{2}\left(1 + \frac{1}{\mu(\mathrm{A})}\right) \tag{17}$$

then the $\mathrm{x}_0$ is the unique solution of the problem.

Now, to express uniqueness theorem for high-dimensional sparse problem, it is sufficient to convert this problem to conventional one using Eq. (11) and (12). Since $\mathrm{A} = \mathrm{A}^{(1)} \otimes \mathrm{A}^{(2)} \ldots \otimes \mathrm{A}^{(D)}$, the *Spark* of the matrix $\mathrm{A}$ is obtained as

$$\mathrm{Spark(A)} \leq \min\left\{\mathrm{Spark}\left(\mathrm{A}^{(1)}\right), \mathrm{Spark}\left(\mathrm{A}^{(2)}\right), \ldots, \mathrm{Spark}\left(\mathrm{A}^{(D)}\right)\right\} \tag{18}$$

So the uniqueness theorem for high-dimensional problem could be expressed as Theorem 3.

**Theorem 3** Consider high-dimensional sparse problem in Eq. (9). The tensor $\mathcal{X}^*$ is a unique solution for it, if and only if

$$\|\mathcal{X}^*\|_0 \leq \frac{1}{2}\min\{\mathrm{Spark}\left(\mathrm{A}^{(1)}\right), \mathrm{Spark}\left(\mathrm{A}^{(2)}\right), \ldots, \mathrm{Spark}\left(\mathrm{A}^{(D)}\right)\} \tag{19}$$

Moreover, the *Coherency* of A could calculated as

$$\mu(\mathrm{A}) = \max\left(\mu\left(\mathrm{A}^{(1)}\right), \mu\left(\mathrm{A}^{(2)}\right), \dots, \mu\left(\mathrm{A}^{(D)}\right)\right) \tag{20}$$

and using this, we can express Theorem 3 in high-dimensional mode as Theorem 4.

**Theorem 4** If the tensor $\mathcal{X}^*$ is a solution to the problem described in Eq. (9) and it satisfies follow condition

$$\|\mathcal{X}^*\|_0 < \frac{1}{2}\left(1 + \frac{1}{\max\left(\mu(\mathrm{A}^{(1)}), \mu(\mathrm{A}^{(2)}), \dots, \mu(\mathrm{A}^{(D)})\right)}\right) \tag{21}$$

then $\mathcal{X}^*$ is an unique solution.

# IV. Proposed Recovery Method

The basis of the proposed method is similar to the SL0 method [15, 20]. In our method, the value of zero-norm is estimated iteratively with a multidimensional Gaussian function. Consider 1-dimensional Gaussian function as

$$f_\sigma\left(x_{j_1 j_2 j_3 \dots j_D}\right) = 1 - e^{-\frac{x_{j_1 j_2 j_3 \dots j_D}{}^2}{2\sigma^2}} \tag{22}$$

If parameter $\sigma$ approaches zero, the value of this function approaches zero, if $x_{j_1 j_2 j_3 \dots j_D}$ is zero and approaches 1 if $x_{j_1 j_2 j_3 \dots j_D}$ is not zero:

$$\lim_{\sigma \to 0} f_\sigma\left(x_{j_1 j_2 j_3 \dots j_D}\right) = \begin{cases} 0, & x_{j_1 j_2 j_3 \dots j_D} = 0 \\ 1, & x_{j_1 j_2 j_3 \dots j_D} \neq 0 \end{cases} \tag{23}$$

This is definition of zero-norm of $x_{j_1 j_2 j_3 \dots j_D}$. So $\|\mathcal{X}\|_0$ could be estimated with function $F_\sigma(\mathcal{X})$ as

$$F_\sigma(\mathcal{X}) = N_1 N_2 \dots N_D - \sum_{j_1=1}^{N_1} \sum_{j_2=1}^{N_2} \dots \sum_{j_D=1}^{N_D} e^{-\frac{x_{j_1 j_2 j_3 \dots j_D}{}^2}{2\sigma^2}} \tag{24}$$

and it can be concluded that if parameter $\sigma$ approaches zero, $F_\sigma(\mathcal{X})$ approaches $\|\mathcal{X}\|_0$. Unlike to zero-norm, the function $F_\sigma(\mathcal{X})$ is convex and differentiable. So gradient optimization algorithms such as steepest decent can be used to minimize it. **Fig. 3**. In this case, the tensor $\mathcal{X}$ has 2 elements $x_1$ and $x_2$. According to the figure, as the value of $\sigma$ decreases, the function $F_\sigma(\mathcal{X})$ behaves closer to the zero-norm.

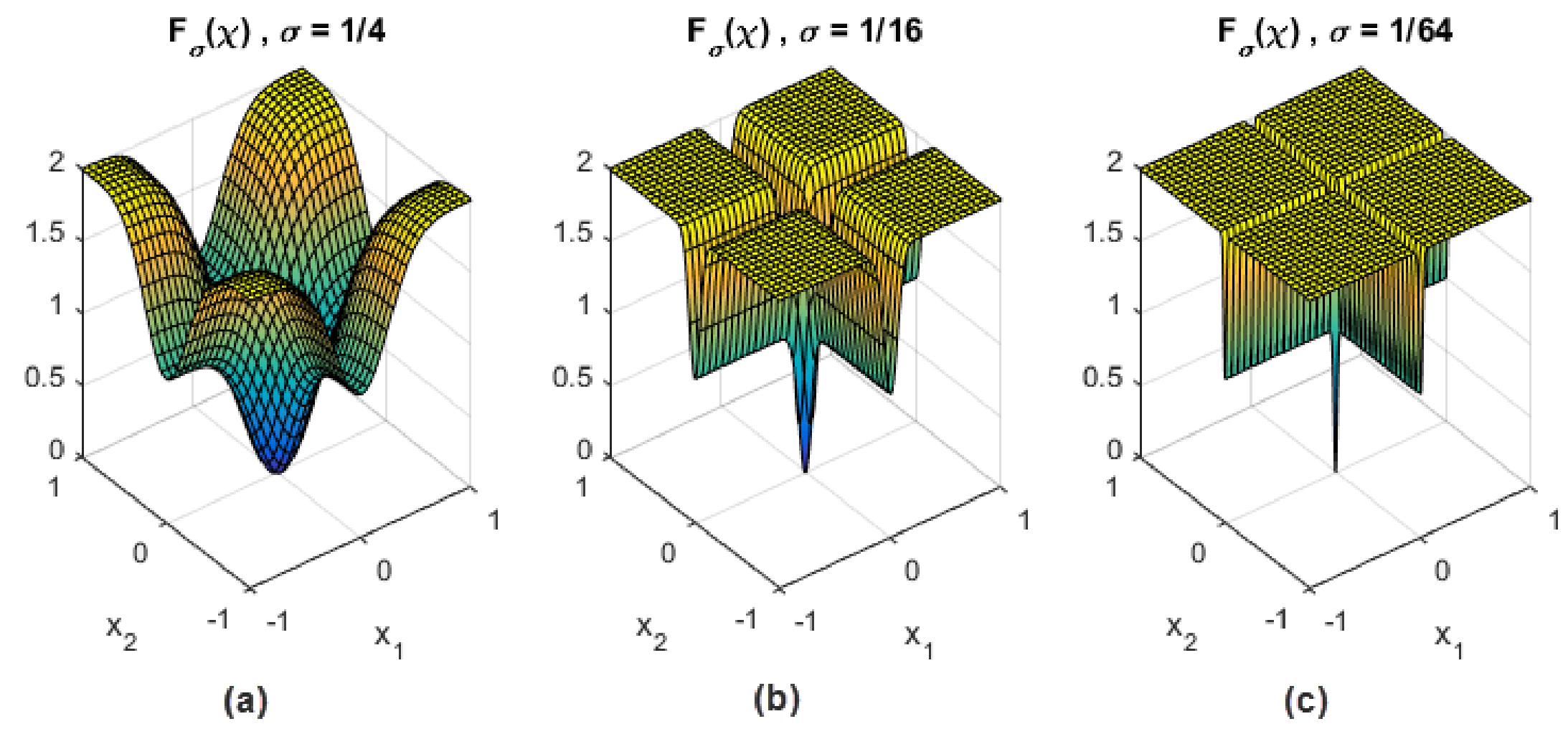

***Fig. 3*** *Displays the function $F_\sigma(\mathcal{X})$ for various values of the parameter $\sigma$.* **(a)** $\sigma = {}^1\!/_4$, **(b)** $\sigma = {}^1\!/_{16}$, **(c)** $\sigma = {}^1\!/_{64}$.

The important point in the cost function $F_\sigma(\mathcal{X})$ is the role of the parameter $\sigma$, which cannot be immediately set to zero or very small value. This is because the function $F_\sigma(\mathcal{X})$ is extremely uneven for near-zero values of $\sigma$, and thus minimizing it will not be easy. The solution is to select a suitable decreasing sequence for $\sigma$ and decrease the value of it step by step. In each step, we will perform an optimization on the function $F_\sigma(\mathcal{X})$ and use its answer as the starting point for the next step. The steepest descent algorithm can be used for optimization.

Another point to be considered is that the solution at each iteration of optimization algorithm should be inside the feasible region. That means the solution should satisfy equation $\mathcal{Y} = \mathcal{X} \times_1 \mathrm{A}^{(1)} \times_2 \mathrm{A}^{(2)} \ldots \times_D \mathrm{A}^{(D)}$ (or inequality $\left\| \mathcal{Y} - \mathcal{X} \times_1 \mathrm{A}^{(1)} \times_2 \mathrm{A}^{(2)} \ldots \times_D \mathrm{A}^{(D)} \right\|_2^2 \leq \varepsilon$ in presence of noise). To do this, the solution in each iteration should be projected on the feasible region. Therefore, the proposed algorithm has 3 major parts. 1. Choosing the initial point for the algorithm. 2. Optimization stage. 3. Projection stage. In the following, we formulate and review these parts in more detail.

1. **Choosing Initial Point:** It has been proved that the answer of steepest descent minimization algorithm, when $\sigma \to \infty$, is equal to the answer of $\ell_2$ (the least square) problem. Hence, a suitable choice for the initial value of the tensor $\mathcal{X}_0$ is equal to the answer of $\ell_2$ problem and is calculated as

$$\mathcal{X}_0 = \mathrm{Y} \times_1 \left(\mathrm{A}^{(1)}\right)^{\dagger} \times_2 \left(\mathrm{A}^{(2)}\right)^{\dagger} \dots \times_D \left(\mathrm{A}^{(D)}\right)^{\dagger} \tag{25}$$

where $\left(\mathrm{A}^{(d)}\right)^{\dagger}$ is pseudo inverse of matrix $\mathrm{A}^{(d)}$ and is calculate as

$$\left(\mathrm{A}^{(d)}\right)^{\dagger} = \left(\mathrm{A}^{(d)}\right)^{\mathrm{T}} \left(\mathrm{A}^{(d)} \left(\mathrm{A}^{(d)}\right)^{\mathrm{T}}\right)^{-1} \tag{26}$$

2. **Optimization Stage:** To minimize the convex function $F_\sigma(\mathcal{X})$, we use steepest descent algorithm, which is a gradient-based method. The update phase of this algorithm for each element of tensor $\mathcal{X}$ is as follows

$$x_{j_1 j_2 j_3 \dots j_D} = x_{j_1 j_2 j_3 \dots j_D} - \mu \frac{\partial F_\sigma}{\partial x_{j_1 j_2 j_3 \dots j_D}} \tag{27}$$

According to definition of $F_\sigma(\mathcal{X})$ in Eq. (24), the partial derivative is calculated as follows

$$\frac{\partial F_\sigma}{\partial x_{j_1 j_2 j_3 \dots j_D}} = \frac{x_{j_1 j_2 j_3 \dots j_D}}{\sigma^2} e^{-\frac{{x_{j_1 j_2 j_3 \dots j_D}}^2}{2\sigma^2}} \tag{28}$$

3. **Projection Stage:** In this stage, the answer minimization of algorithm should be projected on the feasible region. The projection relation for tensor $\mathcal{X}$ is as follows

$$\mathcal{X}_{proj} = \mathcal{X} - \mathcal{R} \times_1 \left(\mathrm{A}^{(1)}\right)^{\dagger} \times_2 \left(\mathrm{A}^{(2)}\right)^{\dagger} \dots \times_D \left(\mathrm{A}^{(D)}\right)^{\dagger} \tag{29}$$

where $\mathcal{R}$ is the remaining tensor, which is the same size as the tensor $\mathcal{Y}$ and is calculated as

$$\mathcal{R} = \mathcal{Y} - \mathcal{X} \times_1 \mathrm{A}^{(1)} \times_2 \mathrm{A}^{(2)} \dots \times_D \mathrm{A}^{(D)} \tag{30}$$

Note that in presence of noise, the projection stage only takes place when the tensor $\mathcal{X}$ obtained from the optimization step does not satisfy inequality $\left\|\mathcal{Y} - \mathcal{X} \times_1 \mathrm{A}^{(1)} \times_2 \mathrm{A}^{(2)} \dots \times_D \mathrm{A}^{(D)}\right\|_2^2 \leq \varepsilon$.

To get the optimal solution, parts 2 and 3 should be repeated to a certain number (e.g., 10 times) inside a loop. Then reduce the value of $\sigma$ and repeat it again. This process should continue until it reaches a specific minimum value for $\sigma$. Finally, the proposed algorithm for sparse recovery in high-dimension can be summarized Algorithm 1.

**Algorithm 1** High-Dimensional Sparse Recovery Using SL0

- Initialization:
  1. Set the initial point of tensor $\mathcal{X}$ as $\mathcal{X}_0 = \mathrm{Y} \times_1 \left(\mathrm{A}^{(1)}\right)^{\dagger} \times_2 \left(\mathrm{A}^{(2)}\right)^{\dagger} \ldots \times_D \left(\mathrm{A}^{(D)}\right)^{\dagger}$.
  2. Choose a suitable descending sequence for $\sigma$ as $[\sigma_1, \sigma_2, \ldots, \sigma_J]$.
- Do following steps for $j = 1, \ldots, J$:
  1. Set $\sigma = \sigma_j$.
  2. Apply steepest descent algorithm on $F_\sigma(\mathcal{X})$ for $L$ times. In each iteration project answer on set $\{\mathcal{X} | \mathcal{Y} = \mathcal{X} \times_1 \mathrm{A}^{(1)} \times_2 \mathrm{A}^{(2)} \ldots \times_D \mathrm{A}^{(D)}\}$:
     - Initial with $\mathcal{X} = \mathcal{X}_{j-1}$
     - Do following steps for $l = 1, \ldots, L$:
       a) Set $\Delta = \left[\delta_{j_1 j_2 j_3 \ldots j_D}\right]$ where $\delta_{j_1 j_2 j_3 \ldots j_D} \triangleq x_{j_1 j_2 j_3 \ldots j_D} e^{-\frac{x_{j_1 j_2 j_3 \ldots j_D}^2}{2\sigma^2}}$.
       b) Update the tensor $\mathcal{X}$ as $\mathcal{X} \leftarrow \mathcal{X} - \mu\Delta$ where $\mu$ is a small positive number.
       c) (In noisy version, if $\left\|\mathcal{Y} - \mathcal{X} \times_1 \mathrm{A}^{(1)} \times_2 \mathrm{A}^{(2)} \ldots \times_D \mathrm{A}^{(D)}\right\|_2^2 > \varepsilon$ then) Project tensor $\mathcal{X}$ on feasible region:
       $$\mathcal{X} \leftarrow \mathcal{X} - \mathcal{R} \times_1 \left(\mathrm{A}^{(1)}\right)^{\dagger} \times_2 \left(\mathrm{A}^{(2)}\right)^{\dagger} \ldots \times_D \left(\mathrm{A}^{(D)}\right)^{\dagger}$$
       where $\mathcal{R} = \mathcal{Y} - \mathcal{X} \times_1 \mathrm{A}^{(1)} \times_2 \mathrm{A}^{(2)} \ldots \times_D \mathrm{A}^{(D)}$.
  3. Set $\mathcal{X}_j = \mathcal{X}$.
- Return final solution as $\mathcal{X}^* = \mathcal{X}_J$.

# V. Numerical Simulation and Discussion

In this section, some numerical simulation are done to evaluate the performance of the proposed method. The parameters of algorithm set as $\sigma_{min} = 0.004$, $L = 5$, $\mu = 0.5$ and the decreasing factor of the parameter $\sigma$ is 0.9 in this simulations. The number of non-zero elements in $\mathcal{X}$ shown with $K$ and the location of them are chosen randomly and their values are generated from a standard normal distribution. Moreover, it is assumed the Gaussian noise with variance $\sigma_n = 10^{-12}$ is added to tensor $\mathcal{X}$ and noise parameter is considered $\varepsilon = 0.01$.

**Simulation 1** In this simulation, we have two example. In first one, $\mathcal{X}$ is considered a sparse matrix whose size is $50 \times 50$ and the size of dictionary matrices $\mathrm{A}^{(1)}, \mathrm{A}^{(2)}$ is $30 \times 50$ which leads to the size of matrix $\mathcal{Y}$ be $50 \times 50$. So the compression will be $1 - \left({}^{30}/_{50}\right)^2 = 0.64$ for this example. Moreover, the number of non-zero samples in $\mathcal{X}$ is equal to $K = 150$. The results are shown in Fig. 4. The original and reconstructed tensors are displayed in Fig. 4 (a) and Fig. 4 (b), respectively. The recovery seems to have been well done. To check precisely, both original and reconstructed matrices are vectorized and their values are plotted in Fig. 4 (c). Two plots are the same and the Signal-to-Noise Ratio for this example is $SNR = 48\ dB$. Note that in our simulations, the initial $SNR$ is $60\ dB$ which is the highest available SNR for recovery algorithms. So achieved SNR is an acceptable value.

The second example is for 3-dimensional tensors. The size of tensor $\mathcal{X}$ is considered $20 \times 20 \times 20$ which 100 elements of it are non-zero ($K = 100$). The size of all three dictionary matrices $\mathrm{A}^{(1)}, \mathrm{A}^{(2)}, \mathrm{A}^{(3)}$ is $12 \times 20$. So, the size of tensor $\mathcal{Y}$ will be $12 \times 12 \times 12$. In this case, compression is equal to $1 - \left({}^{12}/_{20}\right)^3 = 0.784$. The results of this example are shown in Fig. 5.Fig. 5 3-dimensional sparse tensor reconstruction. (a) The exact and **reconstructed** location of non-zero elements in tensor $\mathcal{X}$. **(b)** The exact and recovered values of elements in original and reconstructed tensors(a) demonstrates the exact and reconstructed location of non-zero elements in tensor $\mathcal{X}$. Moreover, the exact and recovered values are shown in Fig. 5 (b). The quality of reconstruction by the proposed algorithm is fine and $SNR = 42\ dB$.

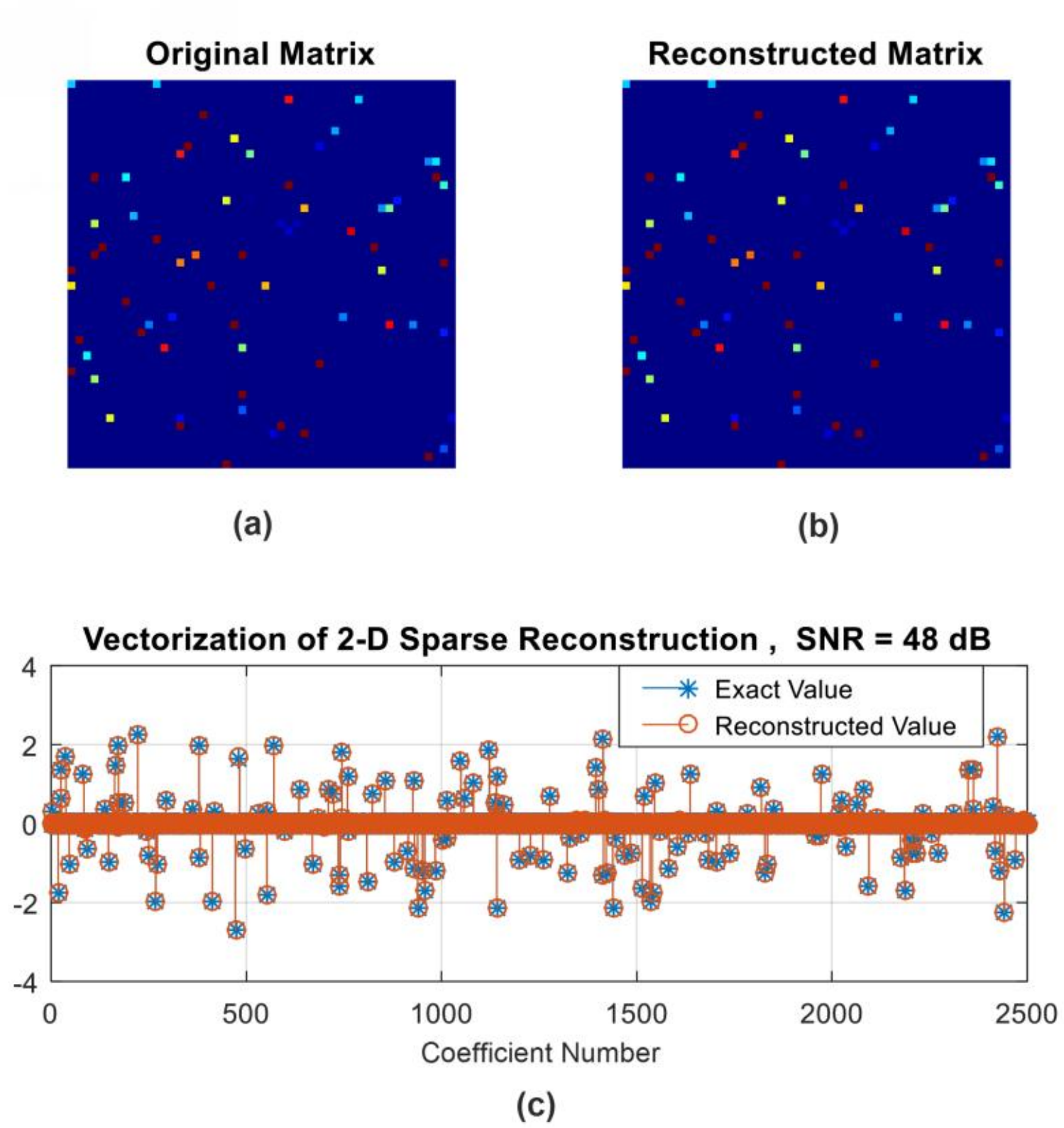

***Fig. 4** 2-dimensional sparse tensor reconstruction.* **(a)** *The original and reconstructed matrices.* **(b)** *The exact and recovered values of matrix elements (Compression = 64%)*

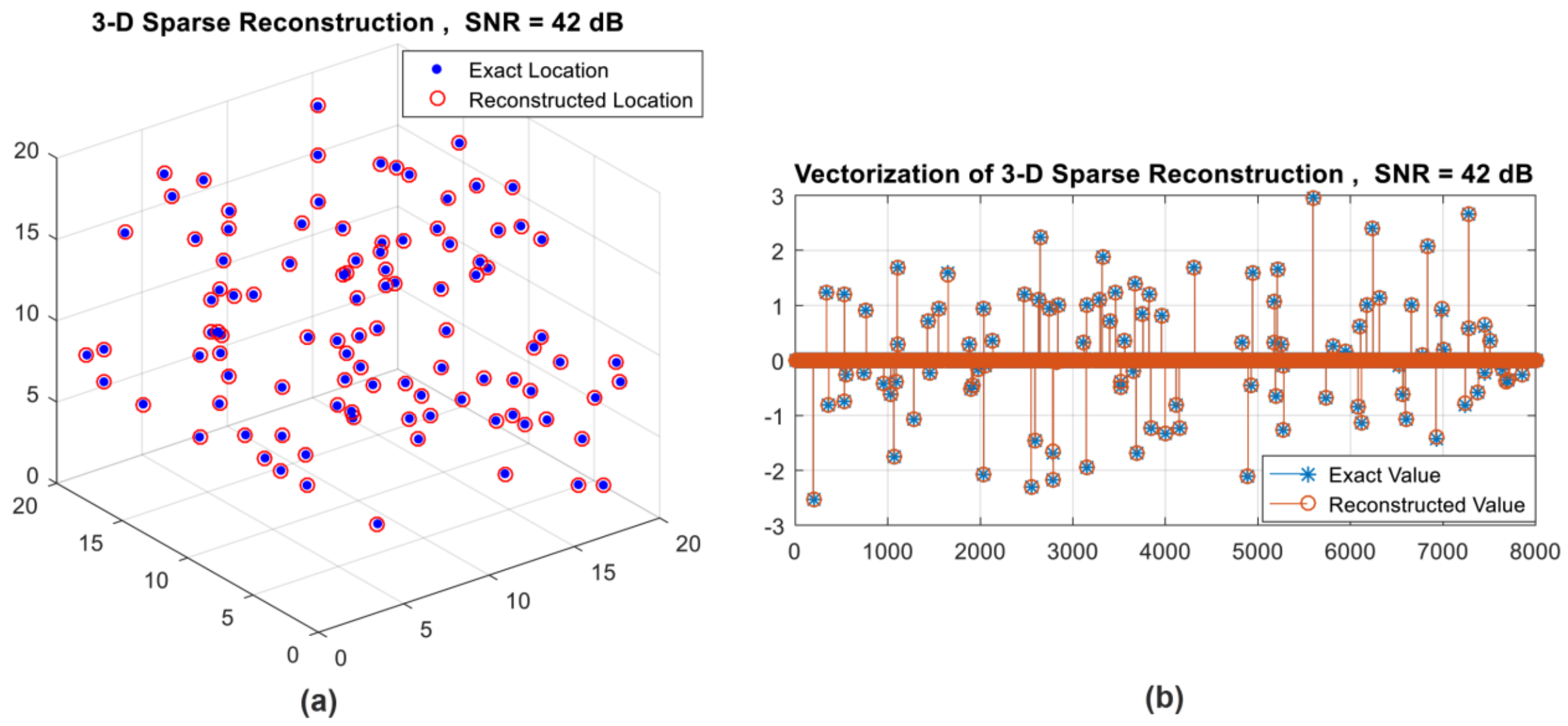

***Fig. 5*** *3-dimensional sparse tensor reconstruction.* **(a)** *The exact and reconstructed location of non-zero elements in tensor $\mathcal{X}$.* **(b)** *The exact and recovered values of elements in original and reconstructed tensors (Compression = 78%)*

**Simulation 2** In second simulation, we try to evaluate the effect of sparsity level of tensor $\mathcal{X}$ on reconstruction quality. For this purpose, $\mathcal{X}$ is considered in 3 dimensions: 1-dimensional, 2-dimensional and 3-dimensional. Fig. 6 displays the reconstruction SNR versus the ratio of the number of non-zero samples in tensor $\mathcal{X}$ to the number of all elements in tensor $\mathcal{Y}$ which shown by $M$. The size of dictionary matrix is $120 \times 200$ for 1-dimensional case and is $12 \times 20$ for 2-dimensioanl and 3-dimensional cases. We increase the number of non-zero samples in $\mathcal{X}$ gradually to examine its effect on SNR. For each case, the simulation is 100 times repeated and the mean SNR in these repetitions is calculated. According to the results, it can be inferred that in 1-dimensional mode, the reconstruction has been performed to a ratio of $K/_M = 0.5$. In two dimensions, this ratio is less and is about 0.25 and is equal to 0.125 in three dimensions.

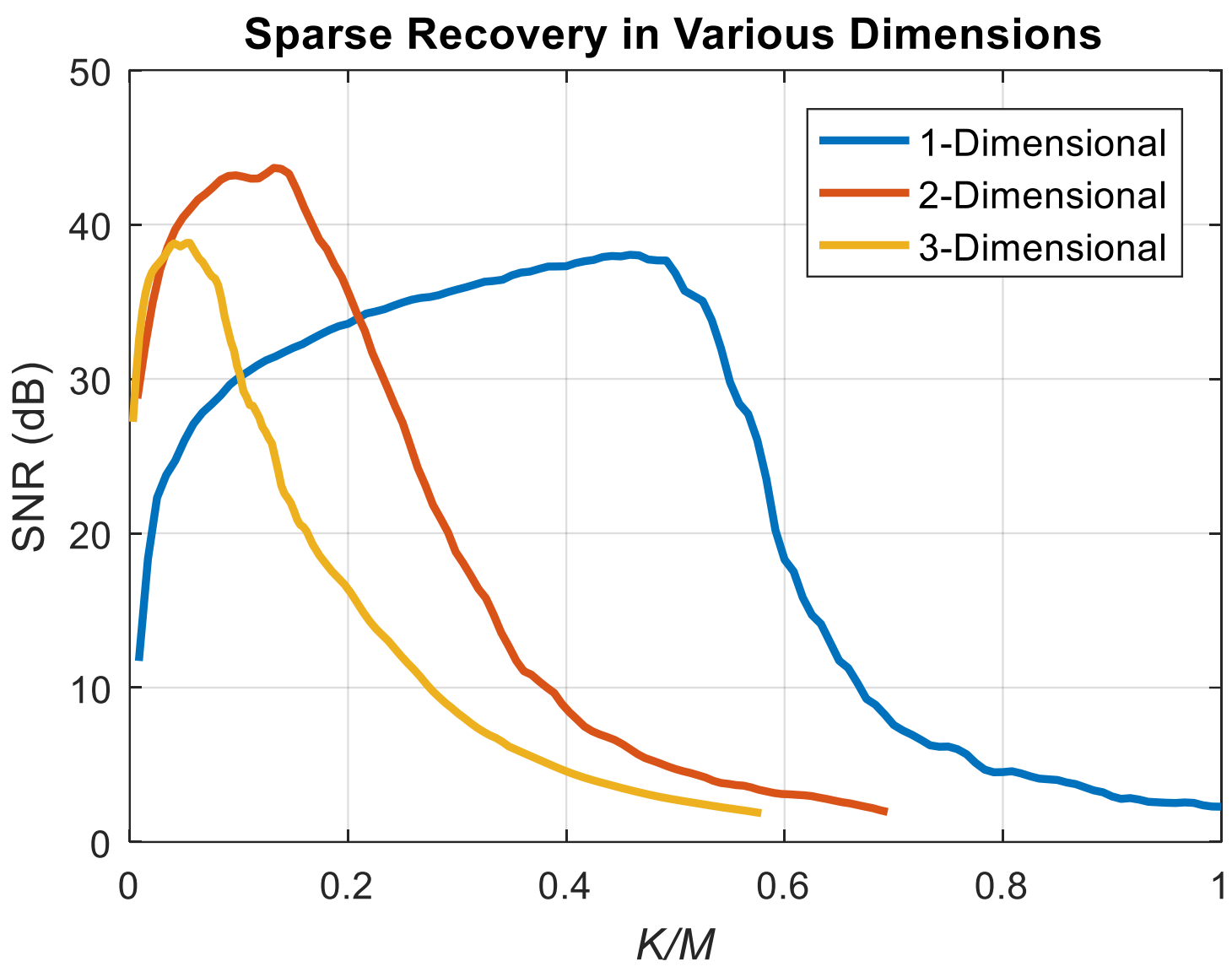

***Fig. 6*** *Sparse Recovery quality in 1-dimension, 2-dimension and 3-dimension versus the proportion* $K/_M$ *where* $K$ *is number of non-zero elements in* $\mathcal{X}$ *and* $M$ *is number of all elements in* $\mathcal{Y}$*.*

Although the upper bound of the sparsity for a unique recovery in high-dimension was expressed as Eq. (19), but it seems for the case where the location of non-zero elements in $\mathcal{X}$ is randomly chosen, the boundary of sparsity in high-dimension could be written as

$$K \leq \left(\frac{1}{2}\right)^D \prod_{d=1}^{D} \mathrm{Spark}\left(\mathrm{A}^{(d)}\right) \tag{31}$$

Since the dictionary matrices are generated randomly, It can be taken that $\mathrm{Spark}\left(\mathrm{A}^{(d)}\right) = m_d$. Considering this assumption, and also assuming that the number of rows in all dictionary matrices is equal to $m$, the right side of Eq. (31) changes as $\left({}^{m}/_{2}\right)^D$. In our simulation, this upper bound is equal to $\left({}^{1}/_{2}\right), \left({}^{1}/_{4}\right), \left({}^{1}/_{8}\right)$ for 1-dimensional, 2-dimensional and 3-dimensional modes, respectively, which is consistent with simulation results. Of course, this bound for uniqueness theorem is only for random mode and is not a definitive guarantee for reconstruction, but according to the simulation results it seems that with high probability reconstruction will be fully accomplished.

**Simulation 3** In the last simulation, we try to examine the main advantage of the proposed algorithm for high-dimensional sparse recovery, which is reduction the time and complexity of the calculations. For this purpose, we consider the three-dimensional tensor $\mathcal{X}$ whose size in each dimension is the same $N_{\mathcal{X}}$. Also, the size of all dictionary matrices is equal to $\left({N_\mathcal{X}}/{2}\right) \times N_\mathcal{X}$ and compression is $(0.5)^3 = 0.875$. The number of non-zeros elements in $\mathcal{X}$ is $K = \left({N_\mathcal{X}}/{5}\right)^3$. Recovery is done in 3-dimensiona. Also, the equivalent 1-dimensional and 2-dimensional problem are formed and solved to comparison complexity in various dimensions. The formation of an equivalent 2-dimensional problem is similar to the 1-dimensional problem. The only difference is that just two of the dictionary matrices are multiplied by the Kronecker product, which means we will have two dictionaries. Also, the tensors $\mathcal{X}$ and $\mathcal{Y}$ are turned to matrix form. Reconstruction is repeated for different values of $N_\mathcal{X}$ and 100 times for each value. Then the average reconstruction time versus the total number of elements in $\mathcal{X}$, which is equal to $N = N_\mathcal{X}^3$, is shown in the Fig. 7. Of course, it should be noted that the SNR of reconstruction in all three method is quite similar and the goal is to compare the recovery time of the methods.

According to the results, it seems that when the number of elements is low, there is no significant difference in time and complexity in three problems. But from Fig. 7 (a) it can be inferred that with increasing number of elements, the reconstruction time of equivalent 1-dimensional problem increases exponentially. So that there are serious problems with time and memory in large sizes, and practically the reconstruction cannot be done. This happens while the 3-dimensional representation problem, even its equivalent 2-dimensional one, is are still very fast up to 10,000 elements.

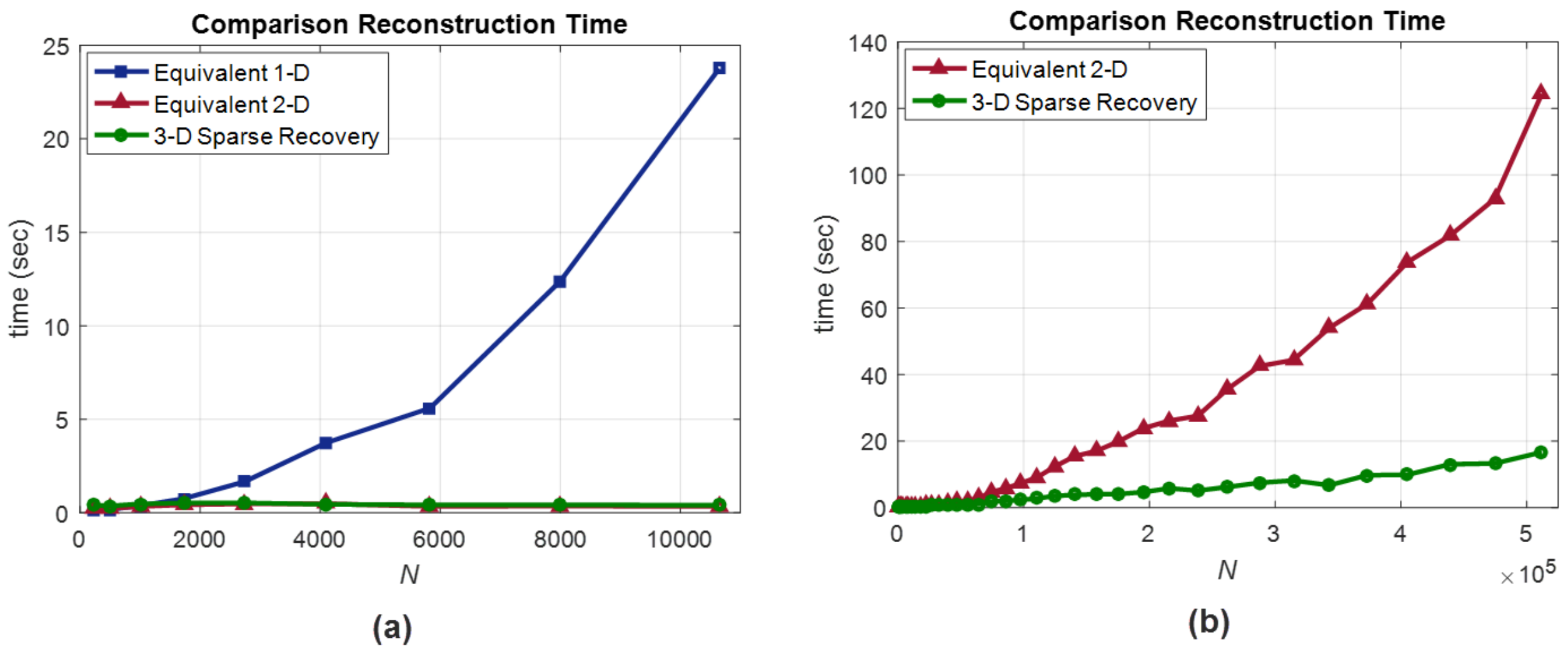

***Fig. 7*** *Comparison reconstruction time between 3-dimensional problem and its equivalent 2-dimenosional and 1- dimensional problems.*

In addition, according to Fig. 7 (b), with more increase in the number of elements, the reconstruction time of 3-dimensional method will be shorter than its 2-dimensional equivalent problem. The more $N$ will increase, the more distance between the two methods will be. Also, in larger sizes, we will face the memory problem in equivalent 2-dimensional problem. The conclusions of this simulation can also be generalized for the higher dimensions.

# VI. Conclusion

In this paper, the sparse representation was generalized in the high-dimensions and formulated in accordance with the theory of tensors. Further, the uniqueness conditions for the solution of the problem in high-dimensions were discussed. Regarding to Section III, it be concluded that for complete reconstruction in high dimensions, the number of non-zero tensor components must be limited to half of the minimum of the dictionary matrices Sparkes. So, it can be inferred that increasing the dimensions lead to more limitation in uniqueness condition.

In the following, a method for solving the high-dimensional sparse representation problem was proposed and its algorithm was developed. At the end, several numerical experiments were performed to validate the proposed method. According to the results, recovery can be achieved by the proposed method completely and with a high equality. Also, the reconstruction time compared to the one-dimensional sparse problem is too short. Of course, it was much better to compare the results of the algorithm with similar algorithms.

The high-dimensional sparse representation problem and the proposed algorithm to solve it can be used in various applications and fields, including MRI imaging, radar imaging, and so on. Using it can cause significant reductions in computing time and improve quality of reconstruction.